\documentclass{eas}
\usepackage{graphicx}
\begin{document}

\TitreGlobal{Mass Profiles and Shapes of Cosmological Structures}

\title{Intracluster stars tracing motions in nearby clusters}
\author{Arnaboldi, M.}\address{INAF, Osservatorio Astronomico di Torino, Strada Osservatorio 20, I-10025 Pino Torinese}
%
\runningtitle{Intracluster Stars}
\setcounter{page}{23}
\index{Arnaboldi, M.}
%
\begin{abstract} 
Cosmological simulations of structure formation predict that galaxies
are dramatically modified by galaxy harassment during the assembly of
galaxy clusters, losing a substantial fraction of their stellar mass
which today must be in the form of intracluster stars.  Simulations
predict non-uniform spatial and radial velocity distributions for
these stars. Intracluster planetary nebulae are the only abundant
component of the intracluster light whose kinematics can be measured
at this time. Comparing these velocity distributions with simulations
will provide a unique opportunity to investigate the hierarchical
cluster formation process as it takes place in the nearby universe.
\end{abstract}

\maketitle

\section{Introduction}
The intracluster light (ICL) in clusters contains a fossil record of
galaxy evolution and interactions in the cluster. It is also relevant
for the baryonic fraction condensed in stars, star formation
efficiency, and the metal enrichment of the hot intracluster (IC) medium
via IC stars, especially in the cluster center.
 
The high resolution simulation of a part of the Universe that
collapses into a galaxy cluster shows that dark matter subhalos grow,
fall into the cluster, may survive or merge into larger halos
(Springel et al. 2001): the same processes may in fact act on stars in
galaxies, producing also ICL. Simulations predict non-uniform spatial
and radial velocity distributions for these stars (Napolitano et
al. 2003) and the intracluster planetary nebulae (ICPNe) are the best
suited tracers for these studies because of their strong [OIII] 5007
\AA\ emission, which allow an easy identification and radial velocity
measurements. Measuring the projected phase-space for the ICPNe allows
us to determine the dynamical age of this component, how and when this
light originated.

\section{ICPNe in the Virgo cluster: projected phase-space distribution}

\subsection{Results from narrow band imaging surveys}
Based on the analysis done by Aguerri et al. (2005), the mean surface
brightness and surface luminosity density of the ICL in several
pointings in the Virgo cluster core fields are $\mu_B = 29.0$ mag
arcsec$^{-2}$ and $2.7 \times 10^6$ L$_{B,\odot}$ arcmin$^{-2}$,
respectively. These values are in good agreement with the
corresponding values obtained from excess red giant counts in two HST
images in the Virgo core. There is no trend evident with distance from
M87. When the fraction of the ICL is computed for the fields in the
Virgo core, it amounts to 5\% of the total galaxy light.

However, the diffuse stellar population in Virgo is inhomogeneous on
scales of $30-90$ arcmin: substantial field-to-field variations are
observed in the number density of PNe and the inferred amount of ICL,
with some fields empty, some fields dominated by extended Virgo galaxy
halos, and some fields dominated by the true IC component.
Furthermore, the qualitative match between the luminosity density
traced by ICPNe and the broadband light detected in the Virgo core by
Mihos et al. (2005) argues that ICPNe are effective tracers of the ICL
in galaxy cluster.

\subsection{Results from spectroscopic surveys}

\begin{figure}
   \centering
\includegraphics[width=6.5cm]{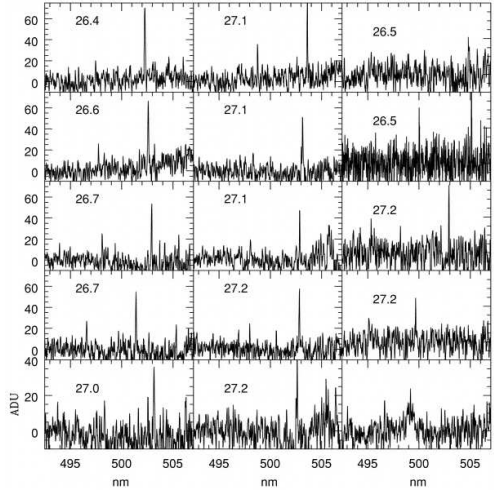}
\caption{FLAMES spectra for 14 ICPNe observed different field
position in the Virgo cluster core. The spectrum in the lower
right corner is a Ly$\alpha$ object, which shows a very broad
line profile. The $m_{(5007)}$ magnitudes are marked on the
individual frames. From Arnaboldi et al. (2004).}
 \label{fig1}
\end{figure}

\begin{figure}
   \centering
\includegraphics[width=6cm]{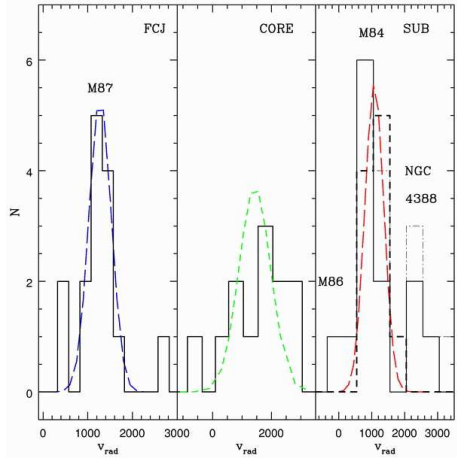}
      \caption{ICPN radial velocity distributions in the three pointings
(FCJ, CORE, and SUB) from Arnaboldi et al. (2004).}
       \label{fig2}
   \end{figure}

Radial velocities of 40 ICPNe in the Virgo cluster were obtained with
the new multifiber FLAMES spectrograph on UT2 at the VLT by Arnaboldi
et al. (2004). For the first time, the $\lambda$ 4959 line of the
[OIII] doublet is seen in a large fraction (50\%) of ICPNe spectra,
see Fig.~\ref{fig1}.  Overall, these velocity measurements confirm the
view that Virgo is a highly non-uniform and unrelaxed galaxy cluster,
consisting of several subunits that have not yet had time to come to
equilibrium in a common gravitational potential, as shown in the
velocity histograms of Fig.~\ref{fig2}.

A well-mixed IC stellar population is seen clearly only in the CORE
field, in the outer parts of the M87 subcluster. Here the velocity
distribution is consistent with a single cluster Gaussian, and the
ICPNe might well have their origin in the tidal effects of the halo of
this subcluster on its galaxy population. In the SUB field near M84
and M86, the ICPNe do not appear virialized; their velocities are
highly correlated with those of the large galaxies in the field. In
fact, there are regions in Virgo where hardly any ICPNe are found, as
in the LPC field of Aguerri et al. (2005). The measurements have also
shown that M87 has a very extended envelope in approximate dynamical
equilibrium, reaching out to at least 65 kpc.

The field-to-field variations, both in number density and in 
velocities, indicate that the ICL is not yet dynamically
mixed. This imposes a constraint on the time of origin of the ICL
and the Virgo cluster itself. The lack of phase mixing suggests
that both have formed in the last few gigayears and that local
processes like galaxy interactions and harassment have played an
important role in this. In a cluster as young and unrelaxed as
Virgo, a substantial fraction of the ICL may still be bound to
the extended halos of galaxies, whereas in denser and older
clusters these halos might already have been stripped.

\section{ICPNe in the Coma  cluster: first detections and future prospects}
\begin{figure}
   \centering
\includegraphics[width=6.5cm]{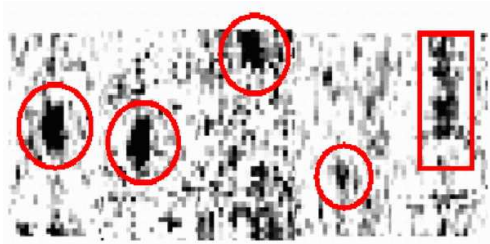}
      \caption{Two-dimensional median-averaged spectra of emission
objects in the MSIS field. The wavelength is along the vertical axis
(507.7-514.15 nm) with a rebinned resolution of 1.5 \AA\ pixel$^{-1}$;
the true spectral resolution is 7.3 \AA\, or 440 km~s$^{-1}$. The
horizontal direction is along the mask slitlets, with a rebinned
resolution of $0''.2$ pixel$^{-1}$. The left four panels show PN
candidates from the brightest to one of the faintest in the field. The
fluxes are (34.7, 31.9, 18.6, 5.6) ADU, corresponding to (17, 16, 9,
3) × 10-19 ergs s$^{-1}$ cm$^{-2}$. The rightmost panel shows the
spectrum of a background galaxy with continuum and strong absorption,
probably blueward of Ly$\alpha$, and a possible line emission,
probably Ly$\alpha$. From Gerhard et al. (2005).}
\label{fig3}
\end{figure}

The Coma cluster is the richest and most compact of the nearby
clusters, where the effect of a dense environment on galaxy evolution
can be studied. Despite being the most compact, there is growing
evidence that its formation is still on-going, and needs to be
investigated further. This is now becoming possible via a multislit
imaging spectroscopy technique (MSIS) using spectrographs on 8 meter
class telescopes.  Gerhard et al. (2005) measured the [O III] $\lambda
5007$ emission lines of 16 ICPN candidates in the Coma cluster with
MSIS and FOCAS, on the 8.2 m Subaru telescope; the two-dimensional
median-averaged spectra of their emission objects in the MSIS field
are shown in Fig.~\ref{fig3}. Comparing with the velocities of Coma
galaxies in the same field, Gerhard et al. (2005) conclude that the
great majority of these candidates would be ICPNe, free floating in
the Coma cluster core. The velocity histogram for the full sample of
ICPN in the MSIS Coma field in now showing presence of substructures
(Arnaboldi et al. in prep.), with the second low velocity peak which
may originate from material of the low-velocity G7 group of galaxies
(Adami et al. 2005).

\end{document}